\documentclass[11pt,twoside]{article}

\usepackage{asp2006}
\usepackage{epsf}
\usepackage{psfig}
\usepackage{lscape}
\usepackage{natbib}

\newcommand{\uv}{ultraviolet}
\newcommand{\ir}{infrared}
\newcommand{\CIV}{C~{\sc iv}}

\newcommand{\OVI}{O~{\sc vi}}
\newcommand{\OVII}{O~{\sc vii}}
\newcommand{\OVIII}{O~{\sc viii}}
\newcommand{\NV}{N~{\sc v}}
\newcommand{\kms}{km~s$^{-1}$}
\newcommand{\msun}{$M_{\odot}$}
\newcommand{\nh}{$N_{\rm H}$}

\begin{document}
\title{Stratified Quasar Winds: Integrating X-ray and Infrared Views of
  Broad Absorption Line Quasars}   
\author{S. C. Gallagher}   
\affil{Department of Physics \& Astronomy, University of California --
  Los Angeles, Los Angeles, CA 90095--1547, USA}    
\author{and J. E. Everett}   
\affil{Departments of Astronomy and Physics, and Center for Magnetic
  Self-Organization, University of Wisconsin -- Madison,
  Madison, WI 53706--1390, USA}    

\begin{abstract} 
Quasars are notable for the luminous power they emit across decades in
frequency from the far-infrared through hard X-rays; emission at
different frequencies emerges from physical scales ranging from AUs to
parsecs.  Each wavelength regime thus offers a different line of sight
into the central engine and a separate probe of outflowing
material. Therefore, obtaining a complete accounting of the physical
characteristics and kinetic power of quasar winds requires a
panchromatic approach.  X-ray and infrared studies are particularly
powerful for covering the range of interesting physical scales and
ionization states of the outflow. We present a stratified wind picture
based on a synthesis of multiwavelength research programs designed to
constrain the nature of mass ejection from radio-quiet quasars.  This
wind comprises three zones: the highly ionized shielding gas, the \uv\
broad absorption line wind, and the cold dusty outflow.  The primary
launching mechanism for the wind likely varies in each zone. While
radiative acceleration on resonance lines dominates for the \uv\
absorbing wind, the shielding gas may instead be driven by magnetic forces.
Ultraviolet continuum radiative pressure, perhaps coupled with magnetic
launching, accelerates a dusty outflow that obscures the inner broad
line region in unification schemes.
\end{abstract}


\section{Introduction}

Theoretical modeling of structure formation in a $\Lambda$-CDM
cosmology cannot match observed galaxy luminosity functions locally
unless some form of heating or ``feedback'' is included in the
simulations \citep[e.g.,][]{GranatoEtal2004}. This strong theoretical
requirement coupled with (1) the empirical discoveries of the strong
correlations between black hole masses and the properties of their
galactic bulges \citep[e.g.,][]{FerMer2000,GebhardtEtal2000} and (2)
the observation that growing supermassive black holes reveal
themselves as luminous quasars \citep[e.g.,][]{Soltan1982,YuTre2002}
have led to quasar winds becoming promising sources of feedback in
massive ($>L^{\ast}$) galaxies \citep[e.g.,][]{SiRe1998}.  Attempts to
model these outflows in cosmological simulations currently employ
simplifying and inaccurate assumptions without incorporating empirical
constraints. Quasar winds are directly observed in Broad Absorption
Line (BAL) quasars; this $\sim15$--20\% of the luminous, radio-quiet
quasar population exhibits deep troughs from high-ionization \uv\
resonance transitions such as \CIV\ and
\OVI. Such absorption features appear blueshifted along lines of sight
passing through winds with terminal velocities reaching 0.03--0.3$c$.

Constraining the nature of quasar outflows, including the geometry,
acceleration mechanism, ionization state, and mass outflow rate, is
fundamental both for understanding the role of quasars in galaxy
evolution as well as accretion physics. Quasars are by nature
multiwavelength, and a complete accounting of the outflow requires a
panchromatic approach.

\section{Empirical Constraints on Wind Properties}

BAL quasars have been the targets of surveys at all wavelengths.  As
the sensitivity of available facilities increases over time, the
empirical data on distinct outflow components have become more
constraining.  Below, we briefly summarize the conclusions and
implications from studies in three regimes, the \uv, the X-ray, and
the \ir; these results are compiled in Table~1.  Constraints on the
launching radius ($R_{\rm launch}$), covering fraction ($f_{\rm
cov}$), column density (\nh), and velocity ($v$) of the wind probed in
these regimes is of particular interest for determining the structure
and kinetic energy of the flow.

\subsection{The Ultraviolet Line-Driven Wind}

BAL quasars by definition show outflows in the \uv, and there is
compelling evidence that these are radiatively driven.  The momentum
from photons absorbed from the quasar continuum is sufficient to push
gas to the high observed velocities of up to $10^{4.5}$~\kms\
\citep[e.g.,][]{MuChGrVo1995,deKool1997}.  
In fact, the BAL absorption features represent photon momentum absorbed from
the \uv\ continuum. In addition, line-locked systems
detected in some objects provide direct evidence for the importance of
line driving
\citep[e.g.,][]{BrMi89}.
Stable absorption-line locking occurs when the relative Doppler shift
of gas at different distances from the continuum source is
approximately equal to the wavelength separation of two strong
absorption lines (e.g., Ly$\alpha$ and \NV). These systems then become
locked into approximately this velocity separation, with the system
closer to the continuum source absorbing the photons that would
otherwise continue to accelerate the more distant, higher velocity
system \citep[e.g.,][]{Scargle1973}.  Line-locked systems do not occur
unless line-driving plays an important role in the dynamics of the
outflow \citep[e.g.,][]{KoVoMoWe1993,Arav1996,ChNe03}.  
In fact, the line-locked transitions themselves have to play an important
role in the dynamics for line-locking to operate.

The geometry most frequently associated with BAL outflows is
equatorial \citep[e.g.,][]{deKBe1995,MuChGrVo1995}, and
spectropolarimetry of BAL and non-BAL~quasars supports this generic
picture \citep[e.g.,][]{OgCoMiTr1999}.  In this scenario, the material
in the wind originates in the accretion disk.  Some vertical pressure,
either thermal, magnetic, or radiative, pushes gas upwards, to be
illuminated by the continuum emission generated at smaller
radii. Radiation pressure then accelerates the gas radially; this is
most efficient when the photon energies match those of strong
resonance atomic transitions.  The covering fraction of the outflow in
this picture is determined by the ratio of the (vertical) disk
pressure to the (radial) central continuum pressure.  As \uv\ emission
lines are often absorbed, the BAL wind must be outside of, or perhaps
co-spatial with, the broad emission line (BEL) region.  Using the
$L_{\rm UV}$--$R_{\rm CIV}$ reverberation mapping relationship
measured by Kaspi et al. (2007)\nocite{KaspiEtal2007}, the \CIV\ BEL
region radius is $\sim2\times10^{17}$~cm for a luminous ($\lambda
L_{1350 \AA}=10^{46}$ erg~s$^{-1}$) quasar.  For a BEL region at the
base of the outflow, the \CIV\ BAL radius is approximately the same
\citep{MC98}.

As mentioned above, because the \uv\ continuum generates the radiative
pressure on the BAL gas, the distance between the continuum and the
BAL gas will ultimately affect the covering fraction of the wind.
This is an area where important observational constraints can perhaps
be brought to bear. Though \mbox{$\alpha$-disk} models predict that $\sim
90\%$ of the optical/\uv\ continuum is generated within $7 \times
10^{15}$~cm for $M_{\rm BH}=3\times10^8$~\msun, recent constraints
from microlensing of quasar accretion disks suggests the continuum
emitting region is actually significantly larger, $\sim 5 \times
10^{16}$~cm \citep{PoEtAl2006,KoEtAl2006}.

\subsection{The Shielding Gas}

At first glance, a quasar appears well-suited to radiative gas
acceleration given the strong \uv\ radiation field.  However, unlike
\mbox{O stars}, quasars are also strong X-ray sources. This high flux
of X-ray photons ionizes the wind, thus eliminating \uv\ resonance
lines.  Highly ionized gas can only be driven radiatively by radiation
pressure on electrons, which is much less efficient than resonance
line pressure.  To prevent overionization, some material is needed to
protect the wind from the ionizing far-\uv\ and X-ray continuum
\citep[][]{ShViSh85}.  In the context of continuous winds, a layer of
shielding gas was hypothesized by Murray et al. (1995; who dubbed it
``hitchhiking gas'')\nocite{MuChGrVo1995} as a thick, highly ionized
layer of gas interior to the \uv\ BAL wind.  Their model required
shielding gas in order to launch the wind from small radii
($\sim10^{16}$ cm).  Though initially introduced in a rather ad hoc
manner, the empirical evidence for the existence of the shielding gas
has become quite compelling.  In particular, measurements of the
column density of X-ray absorbing gas in radio-quiet BAL~quasars find
a range of \nh=10$^{22}$--10$^{24}$~cm$^{2}$
\citep[e.g.,][]{GreenEtal2001,GaBrChGa2002}.  These values are one to
two orders of magnitude larger than the best constraints from careful
modeling of the \uv\ absorption lines \citep[e.g.,][]{AravEtal2001}.
This discrepancy is most dramatic in the BAL quasars whose extreme
X-ray weakness indicates they are likely to host Compton-thick
(\nh\,$>1.5\times10^{24}$~cm$^{-2}$) absorbers.

To date, Compton-thick absorption has only been confirmed for one BAL
quasar, Mrk~231, with the detection of its direct continuum above
10~keV by Braito et al. (2004)\nocite{braito+04}; at softer X-ray
energies only scattered and starburst emission is seen.  Notably, the
putative X-ray Compton-thick BAL~quasars show broad emission lines and
often blue \uv-optical continua 
(e.g., Clavel et al. 2006; Gallagher et al. 2006).\nocite{clavel+06,gall+06}
As first seen by Green et al. (2001)\nocite{GreenEtal2001}, the
$\sim10\%$ of BAL~quasars with low-ionization (Mg~{\sc ii}) BALs may
typically have Compton-thick X-ray absorbers, and Gallagher
et~al. (2006) speculated that Mg~{\sc ii} BALs might require such
X-ray absorption for the low-ionization gas to exist in the outflow.
The converse is not true, however, and so Compton-thick X-ray absorbers
may be a necessary but not sufficient condition for low-ionization
BALs.

Given that an absorber with \nh$>1.5\times10^{24}$~cm$^{-2}$ is
optically thick to the \uv/X-ray continuum in a quasar, such X-ray
absorbers must not fully cover the \uv\ continuum-emitting region.
Though a (less than Compton-thick) highly ionized shielding gas
component might be expected to allow a significant \uv\ photon flux
through the wind \citep{MuChGrVo1995}, radiative transfer calculations
of highly ionized magneto-hydrodynamic (MHD) disk winds hint that some
less ionized gas would still be present to block a large fraction of
the \uv\ flux \citep{Everett05}.  This implies that the \uv\ BAL wind
lies along a distinct path to the \uv\ continuum in these systems
compared to the absorber blocking the X-ray continuum.  If this is
generically true (though only evident for the most extreme examples),
this result provides an important constraint on the relative location
of the \uv\ and X-ray continuum sources.  For an X-ray continuum
generated on smaller spatial scales than the \uv, as implied by recent
constraints from microlensing \citep{PoEtAl2006,KoEtAl2006}, a
stratified wind can account for discrepancies in the \uv\ and X-ray
absorber properties.  We explore this further in \S\ref{sec:model}

To date, the best evidence from X-ray spectral modeling indicates that
the absorbers are plausibly highly ionized such that the soft X-ray
opacity would be dominated by \OVII\ and \OVIII\ absorption edges.
Strong X-ray absorption variability unmatched by changes in the \uv\
BALs also points towards distinct \uv\ and X-ray absorbing material
\citep[e.g.,][]{gall+04} with the X-ray absorber closer than the \uv\
absorber to the central X-ray continuum.  The bulk of the X-ray data
to date thus support the identification of the X-ray absorber with the
putative shielding gas.

\begin{figure}
\centerline{\psfig{figure=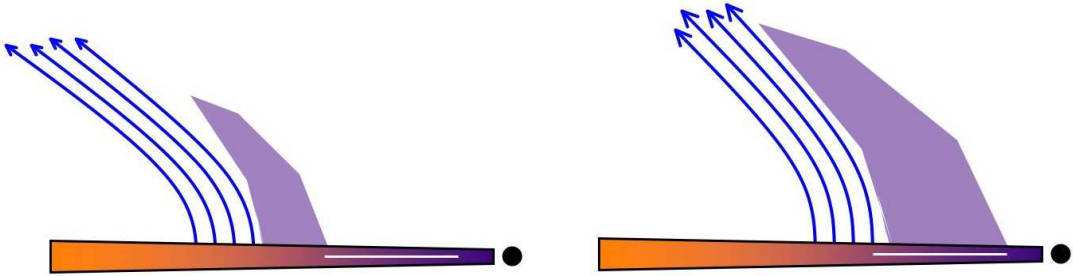,width=15cm}}
\caption{
Two diagrams of the inner part of the accretion disk illustrating the
possible connection between the presence of X-ray shielding gas (solid
shape) and the velocity of the \uv\ BAL outflow (solid curves).  In
both panels, the black hole is to the right, and the continuum
emission is generated in the accretion disk.  The horizontal white bar
serves as a scale marker, and the size of the arrows corresponds to
the velocity of the outflowing BAL gas.  {\em Left:} In this case, a
thin shield does not allow gas to be accelerated until larger radii.
{\em Right:} A thick shield prevents overionization of the \uv\ BAL
wind at smaller radii where the higher photon densities allow the material
to be launched to larger terminal velocities.  A larger vertical
contribution from the accretion disk radiation can also change the
covering fraction of the outflow.
\label{shield}
}
\end{figure}

The recent discovery of a correlation between the maximum terminal
velocity of the \uv\ BAL (as measured for \CIV), $v_{\rm max}$, and
X-ray weakness in BAL quasars indicates the importance of shielding in
the outflow (Gallagher et al. 2006).  Without enough signal in this exploratory
survey for spectral fitting, X-ray weakness was taken
to indicate strong X-ray absorption.  It is a generic property
of gas escaping from the vicinity of the black hole that the terminal
velocity will be of order the Keplerian velocity of the radius from
which it was launched.  Gas that obtains the highest velocities then
might have originated at the smallest radii where the photon densities
are highest.  However, radiative line-driving is only efficient if the
gas does not become overionized; this can be accomplished with a thick
layer of shielding gas.  This scenario might explain the correlation
between extreme X-ray weakness and the highest values of $v_{\rm
max}$.  Figure~\ref{shield} shows a schematic of this situation.

To date, the largest unknown in the properties of the shielding gas is
its velocity.  For the bulk of BAL quasars with X-ray spectra, the
velocity cannot be measured with the current generation of
observatories; only a handful are bright enough in X-rays to search
for X-ray BALs.  Two BAL quasars with putative Fe~{\sc XXV} BALs,
PG1115$+$080 and APM08279$+$5255, indicate high (tenths of $c$)
blueshifts \citep{chartas+07}, however, the frequency of such features
is unknown. 

\subsection{Dusty Outflows}

Outflows may exist on larger scales, as well.  K\"onigl \& Kartje
(1994)\nocite{KoKa1994} first proposed that the so-called ``dusty
torus'' -- the structure consisting of cold material on parsec scales
that reprocesses direct accretion power into thermal \ir\ emission --
is actually an outflow.  In their model (applied to Seyfert galaxies),
the wind is uplifted vertically from the accretion disk along magnetic
field lines as a magneto-centrifugally launched outflow \citep[as in,
  e.g.,][]{BP82,EmBlSh92,Bottorff97}. At large launching radii, dust
can survive in this outflow, and when this dusty gas attains a
sufficient vertical height, it becomes illuminated by the central \uv\
continuum.  At that point, radiation pressure accelerates the dusty
gas radially outward, flattening the wind.  This elegant model avoids
problems with explaining the torus as a large ring of (clumpy) cold
material in the center of the galaxy; such gas is dynamically unstable
and will collapse \citep[c.f.,][]{KrBe1988}.

While the role of magnetic fields in driving quasar outflows remains
to be observationally constrained,
\uv\ photons certainly will efficiently accelerate
dust grains.  The inner wall of this dusty outflow is set by the
temperature at which refractory dust grains (likely graphites)
sublimate, $T_{\rm sub}\sim1500$~K. As the dust is heated by the
radiant power of the quasar continuum, the sublimation radius, $R_{\rm
sub}$, is proportional to $L_{\rm UV}^{0.5}$ \citep{barvainis87}.  For
a quasar with $L_{\rm UV}\sim10^{46}$~erg~s$^{-1}$, the inner wall of
the dusty outflow is at 1--2~pc.  At these radii, the supermassive
black hole will dominate the dynamics of the gas, and the
Keplerian velocity is $\sim10^3$~\kms.  The velocity dispersion of the
bulge of a massive host galaxy, 200--300~\kms, provides a plausible
lower bound to the velocity of this material.

BAL~quasar \uv\ spectra typically show evidence for reddening and
extinction in comparison with non-BAL quasars (Sprayberry \& Foltz 1992;
Reichard et al. 2003)\nocite{SpFo1992,reich+03b}, and this could be
taken to imply that some part of this dusty outflow is perhaps just
the outer regions of the
\uv\ BAL wind.  However,
a recent study of 9.7\micron\ silicate features in \ir\ quasar spectra
by Shi et al. (2006)\nocite{shi+06} found that BAL quasars typically
show very prominent silicate emission.  This is in contrast to type~2
(narrow emission line) Seyfert galaxies which usually show strong
silicate absorption, as predicted by K\"onigl \& Kartje (1994) and
others.  The detection of silicate emission in BAL quasars indicates
that the dusty outflow therefore is distinct from the \uv\ BAL wind.
If the silicate grains (at $T_{\rm sil}\sim200$~K; Hao et
al. 2005)\nocite{HaoEtal2005} were instead carried in the BAL wind,
the line of sight to the \ir\ continuum source generated by the warmer
dust at smaller radii would pass through the silicate region.  In this
case, silicate absorption is expected.

Within the unified quasar picture, some fraction of the sky is
obscured by the dusty outflow such that the broad emission line region
and central continuum are hidden from the direct line of sight.  In
this case, the ratio of type~2 to type~1 (broad emission line) quasars
gives the value for the covering fraction of the dusty outflow
\citep[e.g.,][]{ric+06}.\\

\begin{centering}
\begin{tabular}{lccccc}
\multicolumn{6}{c}{Properties of the Stratified Quasar Wind}\\
\hline
Wind           &  $R_{\rm launch}$        & $f_{\rm cov}$      & \nh$^{\rm a}$ &
ion. & $v$            \\
Component      &  (cm)       &                    & (cm$^{-2}$)   &
state$^{\rm b}$  &(\kms)  \\   
\hline
Shielding Gas  & 10$^{15-16}$   & $>f_{\rm cov,UV}$  & 10$^{22-24}$  &
\OVII, \OVIII\ &?             \\ 
UV BAL Wind        & 10$^{17}$   & 0.2(1--$f_{\rm type2}$) &  10$^{21-22}$
& \CIV, \OVI\ & 10$^{3-4}$ \\
Dusty Outflow  & 10$^{18.5}$ & $f_{\rm type2}$     &   $\cdots$   &
neutral & 10$^{2-3}$ \\
\hline
\end{tabular}

$^{\rm a}$Line-of-sight column density. $^{\rm b}$Common ions
representing the ionization state.
\end{centering}

\section{The Role of Winds}

In the simplest disk-wind paradigm, all quasars host outflows, but
only in BAL~quasars are these driven along the line of sight.
Therefore, the fraction of type~1 quasars with BALs, $\sim15$--20\%
\citep{reich+03b,HewFol2003}, corresponds to the covering fraction of
the BAL wind.  The opposite case would be that only a subset of
quasars host BAL winds, but these dusty shrouds cover a large fraction
of the sky -- the ``cocoon'' picture \citep[e.g.,][]{BeckerEtal2000}.
In this latter situation, BAL quasars would be expected to be mid-\ir\
bright relative to non-BAL~quasars with little or no wind because a
larger fraction of the accretion power is captured and reprocessed
into the thermal \ir\ by dust.  The recent {\em Spitzer} survey of 38
BAL~quasars by Gallagher et al. (2007)\nocite{gall+06c} disputes this
latter view, as they found that the mid-\ir\ properties of BAL~quasars
are consistent with non-BAL~quasars of comparable luminosity.  In
particular, the relative power in the optical and mid-\ir\ in the two
populations is indistinguishable.  Coupled with clear evidence from
spectropolarimetry that there are lines of sight to BAL quasars that
are not covered by the \uv\ outflow \citep[e.g.,][]{OgCoMiTr1999}, it
seems quite likely that most luminous quasars host BAL outflows, and
only in BAL~quasars are we actually looking through them
\citep{WeMoFoHe1991}.

Though the first \uv\ spectroscopic comparisons of the emission-line
and continuum properties of BAL versus non-BAL quasars found them to
be ``remarkably similar'' \citep{WeMoFoHe1991}, spectral studies with
much larger samples revealed that BAL are more often found in quasars
with intrinsically blue \uv-optical continua and broader emission lines
\citep{ric+02}.  We point out that the covering
fraction of the wind in any given quasar is likely to vary, and so
those quasars with the largest covering fractions are most likely to
be identified as BAL quasars.  This will skew the ``average''
continuum and emission-line properties of BAL quasars to be
representative of quasars with more substantial outflows, rather than
the typical outflow.  A discussion of possible links between active
winds, continuum properties, and \uv\ emission-line properties is
presented in Richards (2006).\nocite{ric06}

\section{Constructing a Consistent Geometry}
\label{sec:model}

Based on the empirical constraints outlined in \S2, we construct the
diagram of the stratified wind presented in Figure~2 with
approximately three zones: the shielding gas, the \uv\ BAL wind, and
the dusty outflow.  Each zone is spatially distinct and can be
characterized by distinct covering fractions, column densities,
ionization states, and probably velocities.  The acceleration
mechanisms also differ. While there is compelling evidence that
radiative line pressure dominates for the \uv\ BAL wind, the dynamical
state of the shielding gas remains uncertain.  For gas characterized
by atomic species from
\OVII\ up to Fe~{\sc xxv}, the gas is too ionized for line driving to
be effective, and X-ray continuum driving is also insufficient
\citep{EvBa2004}.  Therefore, if the shielding gas velocities are
typically $\sim0.1c$ as seen in the two known cases of X-ray BALs, MHD
forces are likely to dominate the acceleration.  However, the
shielding gas might instead be stalled or even infalling
\citep{PrStKa2000}.  Meanwhile, for the dusty outflows on large
scales, \uv\ continuum pressure on dust grains overrides electron
continuum pressure by approximately a factor of 850 \citep{KoKa1994}.
Efficient acceleration combined with the large launching radius for
the dusty component make it the most equatorial part of the outflow in
a luminous quasar.

Observationally, panchromatic observations are very important in
constructing this picture, as each wind component is viewed primarily
in a distinct portion of the spectral energy distribution.  While the
X-ray continuum is imprinted by both the shielding gas and the \uv\
BAL wind, the larger column density of the shielding gas makes its
effect more pronounced.  The \uv\ continuum is likely not completely
covered by the shielding gas, which in any case is highly ionized and
would be nearly invisible in the \uv\ for \nh$\ll 10^{24}$~cm$^{-2}$.
The BALs affecting the \uv\ continuum are the clear signatures of this
component.  The dusty outflow, meanwhile, is detected and probed via
its \ir\ emission in luminous BAL quasars.

We emphasize that this proposed picture is based on empirical data for
luminous, radio-quiet BAL quasars. Outflows in both lower luminosity
Seyfert galaxies and radio-loud quasars are likely to be qualitatively
distinct because of differences in both luminosity and spectral energy
distributions.  Specifically, Seyfert galaxies and radio-loud quasars
emit a larger fraction of their radiant power in the X-rays than
luminous radio-quiet quasars \citep[e.g.,][]{BrYuSi1997,steffen+06}.
As discussed by Murray et al. (1995), X-ray loud active galactic
nuclei will have difficulty launching radiatively driven winds.  
These differences in wind driving are observationally supported: for
instance, the \uv\ absorbing outflows in the lower luminosity Seyfert
galaxy NGC 4151 are likely MHD-driven \citep{CrKr2006}.  In addition,
magneto-centrifugally dominated wind models have successfully fit
emission line variations in the Seyfert 1 galaxy NGC 5548
\citep{Bottorff97}; these models also yield a stratified wind
structure \citep{BoKoSh2000}.

\begin{figure}
\psfig{figure=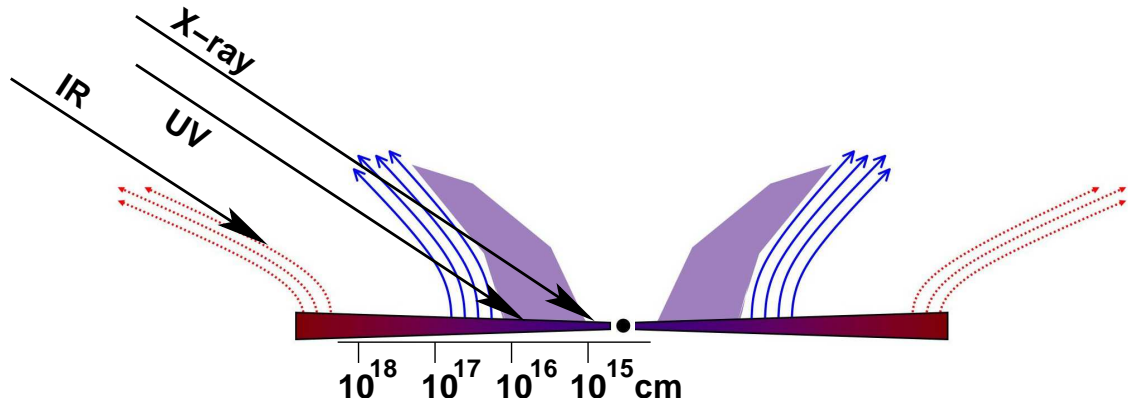,width=15cm}
\caption{
A diagram of the black hole and accretion disk of a luminous,
radio-quiet quasar illustrating the separate components of the outflow
as described in the text: the shielding gas (solid shapes), the \uv\
BAL wind (solid curves), and the dusty outflow (dotted curves).  The
straight, labeled lines indicate the observer's lines of sight through
the stratified wind probed by studies in the X-ray, \uv, and \ir;
these arrows point to the approximate location of the continuum source
in each regime.
\label{model}
}
\end{figure}

\section{Conclusions}

Figure~2 is a schematic, simplified view of the stratified wind drawn
in an attempt to incorporate the growing body of multiwavelength data
as well as modeling of disk winds from the past decade or so.  As
such, it requires further theoretical and empirical elaboration.  For
example, the outflowing (magneto-centrifugal) wind is likely to be
clumpy, as suggested by comparisons of models with observations
\citep{NeIvEl2002,ElSh2006}.  Furthermore, the shielding gas is
probably not discontinuous from the \uv\ BAL wind, but is instead the
highly ionized inner region
\citep[e.g.,][]{KoKa1994,MuChGrVo1995,BoKoSh2000,PrStKa2000,Everett05}.

Outflows are most easily studied in BAL quasars where the absorbing
gas is obviously along the line of sight.  However, there should be
signatures of outflows in non-BAL quasars if the disk-wind paradigm is
generally correct.  For example, X-rays absorbed by the shielding gas
will be emitted along other lines of sight; this contribution to
non-BAL quasar X-ray spectra at soft energies will depend on the
covering fraction and geometry of the shield.  In this case, high
signal-to-noise X-ray spectroscopy may reveal a variable scattered
light component in luminous type~1 quasars.

In the near future, high quality near and mid-\ir\ spectra will offer
new insights into the hottest dust at the inner boundary of the dusty
outflow.  Furthermore, in-depth analysis of solid state features such
as 9.7 and 18~\micron\ silicate emission can provide constraints on
dust processing and perhaps grain formation within the quasar
environment.  

At present, it appears that neither the dusty outflow nor the \uv\ BAL
wind carries sufficient kinetic luminosity to account for the feedback
required to affect galaxy evolution.  The shielding gas, with its high
column density and currently unknown velocity, is therefore the most
promising component to dominate the energetics. Constraining these
velocities will require the high spectral resolution and large
effective area of the next generation of X-ray observatories such as
{\em Constellation-X}, as well as continued modeling efforts that
incorporate all phases of the outflow.


\acknowledgements 
We thank M. Elitzur,  F. Hamann, D. Hines, A. K\"{o}nigl, and
G. Richards for helpful discussions that contributed to the picture
presented in this paper, and O. Blaes and S. Kaspi for pointing us
towards useful empirical results.  Support for S.C.G. was provided
by NASA through the {\em Spitzer} Fellowship Program, under award
1256317; this project was made possible by {\em Chandra} X-Ray Center
grant GO4--5113X. J.E.E was supported by NSF AST--0507367 and NSF
PHY--0215581 (to the Center for Magnetic Self-Organization in
Laboratory and Astrophysical Plasmas).



\end{document}